\documentclass[aps,pra,reprint,superscriptaddress,amsmath,amssymb]{revtex4-2}
\usepackage{graphicx}
\usepackage{xcolor}
\usepackage[colorlinks=true,bookmarks=false,citecolor=blue,urlcolor=blue]{hyperref} 
\usepackage{float}

\usepackage{orcidlink}
\begin{document}
\title{Robust bistable vortex light bullets in graded-index multimode fibers  }
\author{Ashis Paul,\orcidlink{0000-0002-9656-9298}}
\affiliation{Department of Physical and Chemical Sciences, University of L'Aquila, Via Vetoio, 67100 L'Aquila, Italy}
\author{Anuj P. Lara,\orcidlink{0000-0002-8042-7754}}
\affiliation{Department of Applied Physics and Electronics, Umeå University, Umeå 90187, Sweden}
\author{Govind P. Agrawal}
\affiliation{The Institute of Optics, University of Rochester, Rochester, New York 14627, USA}
\author{Samudra Roy,\orcidlink{0000-0001-6178-5516}}
\affiliation{Department of Physics, Indian Institute of Technology Kharagpur, West Bengal 721302, India}

\begin{abstract}

We explore theoretically the formation and evolution of spatiotemporal vortex
bullets, pulses confined in both space and time while carrying orbital angular
momentum, inside a graded-index multimode fiber. Through a two-fold approach,
combining variational analysis with full numerical simulations, we establish
the existence of a bistable vortex soliton family, characterized by distinct
radial and azimuthal quantum numbers. These wave packets exhibit a
Laguerre-Gaussian multi-ring topology in their spatiotemporal structure. We
identify the experimental conditions necessary for the generation of such
structured states of light. A stability analysis based on Vakhitov-Kolokolov
criteria reveals an upper limit for the propagation constant, below which
vortex bullets are found to be stable. Stationary and non-stationary dynamics
of vortex bullets with different topological charges are analyzed fully,
exploiting analytical techniques that are supported by full numerical
simulations. These results enhance our understanding of self-trapped,
spatiotemporal vortex solitons in a graded-index medium and may lead to
potential applications in the area of classical and quantum information
processing.

\end{abstract}
\maketitle

\section{Introduction}

Optical vortices are characterized by a helical phase structure with an on-axis
phase singularity and a zero intensity at the beam's center
\cite{Desyatnikov05, Malomedrev}. Mathematically, their electric field contains
a phase term of the form $e^{il\phi}$, where $\phi$ is the azimuthal angle in
cylindrical coordinates. The integer $l$ is called topological charge and is a
measure of the orbital angular momentum (OAM) associated with an optical
vortex. The most well-known example of such vortices are continuous-wave (CW)
beams with a Laguerre-Gaussian (LG$_{pl}$) spatial profile characterized by two
positive integers $l$ and $p$. Such beams carry an OAM of $l\hbar$ per photon
\cite{Allen92,SkryabinOAM}, and they play a key role in diverse areas of
physics, including optical trapping \cite{Ashkin87,simpson97}, OAM-controlled
particle manipulation \cite{dholakia2011,padgett2011,shen19}, superresolution
microscopy \cite{Spektor08}, free-space communication
\cite{wang2012,Bozinovic13,wilner15}, holographic data storage\cite{Kong23},
quantum computing \cite{Vallone2004,Mafu2013, Nicolas14, erhad2018,Forbes19},
and encryption~\cite{Fang2019}.

In recent years, spatiotemporal optical vortices (STOVs), pulses confined in
both space and time, have attracted attention due to their ability to carry
transverse OAM \cite{Aiello15}. Spurred by several theoretical proposals
\cite{Denisenko09,Biokh2012,Porras2019,hancockprl2021,fangprl21,ChenAcs22} and
recent breakthroughs in ultrafast optics, STOVs have been realized in diverse
experimental settings \cite{prx2016,hancock2019,chongnature2020,
zhao20,guinat2021,LiuLG25,Huo24,Huang24,Liu2024,Fan2025}. In contrast to CW
vortices, STOVs possess phase circulation in the space-time domain, which has
opened up several key applications~\cite{shen19,Aiello15}.

The formation of stationary STOVs, called STOV bullets, has been of interest,
following the discovery of spatial and temporal solitons in a Kerr medium
\cite{kivshar2003}. In fact, since its initial proposal \cite{silberberg1990},
such spatiotemporal wave packets are referred to as light bullets (LBs). As
their formation results from a delicate balance between dispersion,
diffraction, and the nonlinearity, they have remained an active area of
research \cite{malomedrev2005,malomedbook}. It is known that LBs are unstable
in materials exhibiting self-focusing Kerr nonlinearity, where the beam
undergoes a supercritical collapse after a finite propagation distance. Among
discrete systems, evanescently coupled waveguide arrays have emerged as a
promising candidate to support stable LBs \cite{aceves1994,Milan2019}, leading
to experiments on LBs \cite{minardi2010} and their vortex counterpart
\cite{eilenberger2013}. Among continuous systems, several mechanisms have been
proposed over the years to stabilize LBs, including saturable absorption
\cite{skarka1997,veretenov2016,veretenov2017}, photonic lattices
\cite{mihalache2004,mihalache2005,kartashovPT2016,kartashovml2022}, competing
\cite{mihalache2002,mihalacheprl2006,skarkaprl2006,mihalache2007} and nonlocal
\cite{obang2002,mihalacheNL2006,burgess2009} nonlinearities, and others
\cite{zhang2015,dong2021,ivanov2023,java2016,gopala2021,sun2023,driben2014,kartashov2014,Paul2024}.

Recently, spatial modulation of the refractive index has been shown to
stabilize a LB through the waveguiding effect \cite{Malomed26}. Graded-index
(GRIN) multimode fibers (MMFs) provide an example of optical waveguides in
which refractive index varies in a parabolic fashion, and they should support
the LBs proposed in Ref.~\cite{Malomed26}. The open question is whether GRIN
fibers can also provide a stabilization mechanism for STOV bullets. Our
analysis in this paper shows that the answer is affirmative. MMFs have
attracted renewed interest for enhancing the capacity of telecommunication
systems through mode-division multiplexing \cite{RichardsonSDM}. However,
pioneering work during the 1980s \cite{Hasegawa80,Crosignani81,Grudinin88}
predicted the existence of MMF solitons, followed by several studies focusing
on a periodic self-imaging effect in a parabolic GRIN media
\cite{Karlsson1992,Longhi2004,WrightNature2015}. Multimode solitons, composed
of a few spatial modes, have been observed using a GRIN fiber
\cite{WiseNature2013,Zhu2016}. Variational analysis has also been used to show
that LB-type solutions ($l=0$, no OAM) are possible in GRIN media \cite{Yu1995,
GPA_raghaban_2000,TuritsynPRA18,Ahsan2018,ParraRivasLB23}. A recent study has
shown that higher-order dispersive effects can also stabilize such LBs
\cite{PRivasQB2024}. Spatiotemporal vortex LBs with nonzero vorticity ($l>0$ remain of
continued interest \cite{KrupaMMFrev19,Sun2024,su2025,Delgado25,Guo21,dong24}.

In this article, we show that bistable STOV bullets can form in silica-gased
GRIN fibers. We use a two-fold approach, combining variational analysis with
full numerical simulations based on the split-step Fourier method
\cite{GPA_NLbook}. In Section II, we adopt a simpler nonmodal approach
\cite{Conforti2017} and present the (3+1)D nonlinear Schr\"{o}dinger equation
(NLSE) used to study the evolution of STOV bullets inside a GRIN fiber. In
Section III, we apply the variational analysis and obtain bistable solutions
representing a family of STOV bullets with different energies. In Section IV,
we examine the stability of these solutions through the Vakhitov--Kolokolov
(VK) criterion \cite{VKpaper} and compare its prediction with full numerical
simulations. Section V is devoted to investigate the stability of STOV bullets
under perturbations. We summarize our work in Section VI.

\begin{figure}[tb!]
    \centering\includegraphics[width=0.5\textwidth]{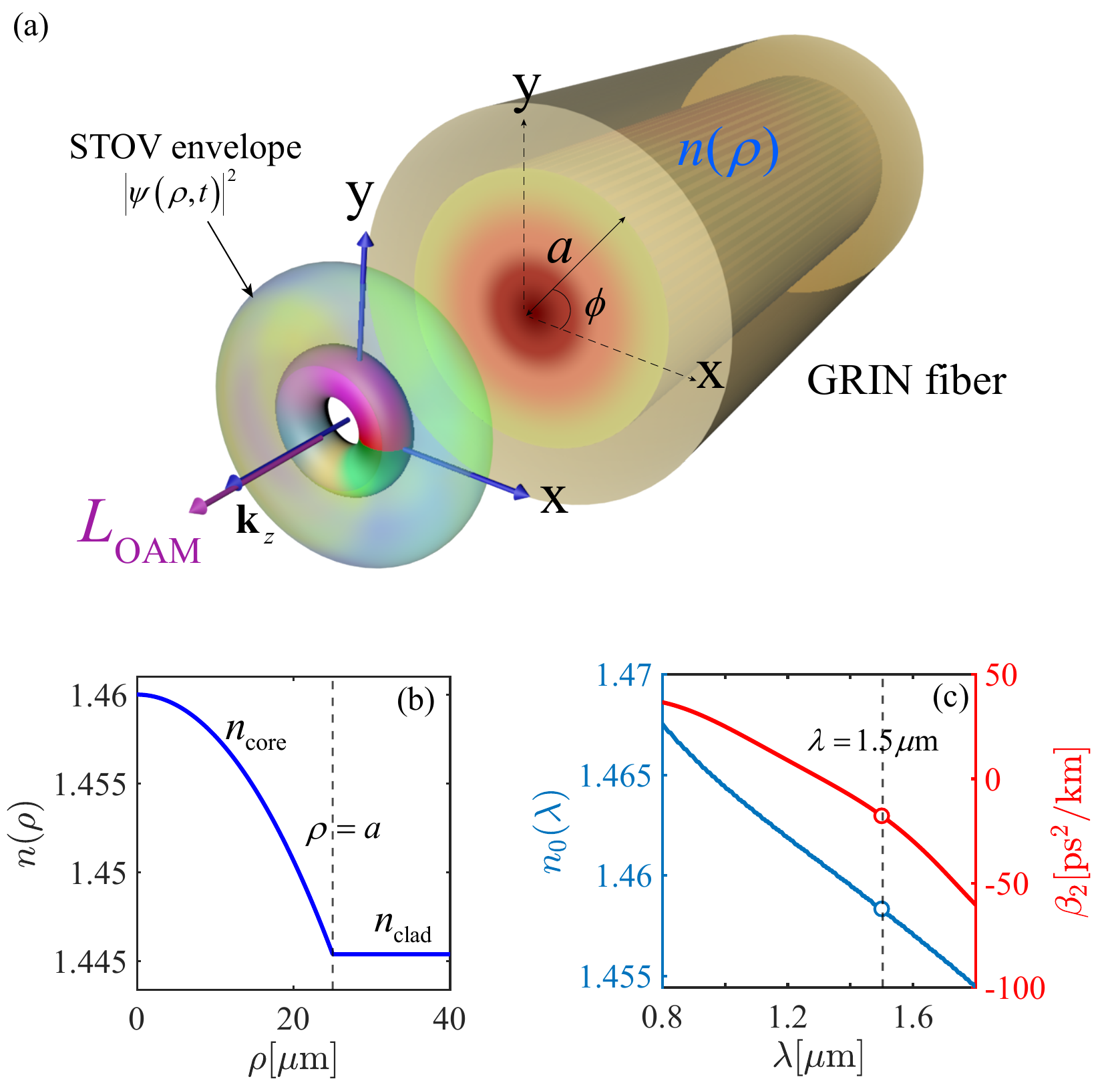}
\caption{(a) Schematic illustration of a radially symmetric STOV bullet inside a GRIN fiber. (b) Refractive index profile for a GRIN fiber with $a=25~\mu$m and $\Delta=0.01$. A dashed line shows the core-cladding boundary. (c) Material dispersion $n_0(\lambda)$ (blue curve) and $\beta_2(\lambda)$ (red curve) for the GRIN fiber. The vertical dashed line at $\lambda =1.5~\mu$m marks the input wavelength, where  GVD is $\beta_2 = -17\; \rm ps^2/km$. The GVD is calculated for a GeO$_{2}$ mole fraction of 0.0862 \cite{GeO2}.} \label{Fig1}
\end{figure}

\section{(3+1)D nonlinear Schr\"{o}dinger equation}

We consider a GeO$_2$-doped silica GRIN fiber whose refractive index varies parabolically inside the core as
\begin{equation}
    n(\textbf{r},I,\omega) = n_{0}(\omega) \bigg[1-\frac{1}{2}b^2 \rho^2\bigg] + n_2 I(\textbf{r}), \quad \; \rho \le a
    \label{RI}
\end{equation}
where $\rho = \sqrt{x^2+y^2}$ is the radial distance from the central axis of the GRIN fiber, $n_{0}(\omega)$ is the value at the center $\rho =0$, $n_2=2.7\times 10^{-20} \rm m^2/W$ is the Kerr coefficient, and $I$ is the local intensity. The index gradient $b$ is defined as $b = \sqrt{2\Delta }/a$, where $a=25\;\mu$m is the core radius and  $\Delta =(n_{0}-n_{\rm cl})/n_{0}$, where $n_{\rm cl}$ is the cladding index. Figure \ref{Fig1}(a) shows schematically a radially symmetric STOV bullet forming inside such a GRIN fiber. For this work, we use $\Delta = 0.01$, which can be realized through GeO$_2$ doping with mole fraction of $0.0862$ \cite{GeO2}. This choice yields $ b=5.7 \;\rm mm^{-1}$ for which the parabolic index profile is depicted in Fig.~\ref{Fig1}(b).

We consider linearly polarized pulses with the electric field in the form
\begin{equation}\label{gpa1}
    \textbf{E}(\textbf{r},t)=\hat{\textbf{p}} \text{Re} \big[E(\textbf{r},t)e^{i \beta_0 z - i \omega_0 t  }\big],
\end{equation}
where $E(\textbf{r},t)$ is the slowly varying field amplitude, localized both in space and time, and $\hat{\textbf{p}}$ is the unit vector representing the state of polarization of the beam. In the following, we use $\hat{\textbf{p}}= \hat{\textbf{x}}$. The dispersion relation is $\beta(\omega)= \omega n(\omega) /c$, and $\beta_0 = \beta(\omega_0)$ is its magnitude at the carrier frequency $\omega_0$. In the paraxial and slowly varying envelop approximations, the (3+1)D NLSE governing the spatiotemporal dynamics in the parabolic GRIN medium has the form~\cite{GPA_grinbook}
\begin{eqnarray} \label{PropEq1}
    i\frac{\partial E }{\partial z} + \frac{1}{2\beta_0}\nabla^2_{\bot} E + \hat{\mathcal{D}} \left(i\frac{\partial}{\partial \tau}\right)E - \beta_0 \Delta\frac{\rho^2}{a^2} E \nonumber\\
    + \frac{n_2\omega_0}{c A_{\rm eff}} |E|^2 E =0,
\end{eqnarray}
where $\nabla^2_{\bot}=\partial^2/\partial x^2 + \partial^2/\partial y^2$ is the transverse Laplacian operator, $\tau=t- z/v_g$ is the local time in the comoving pulse frame, and $v_g=\beta_1^{-1}$ is the group velocity. The effective mode area $A_{\rm eff}$ depends on the spatial distribution of the fundamental mode. By approximating it with a Gaussian distribution, one obtains $A_{\rm eff}=\pi w_g^2$, where $w_g=1/\sqrt{\beta_0 b}$ is the width of the fundamental mode of the GRIN fiber in the absence of Kerr nonlinearity.
The operator $\hat{\mathcal{D}}$ in Eq.\ \ref{PropEq1} has the following form in the frequency domain:
\begin{equation}\label{gpa2}
    \hat{\mathcal{D}}(\omega)=\sum_{m=2}^{\infty} \beta_m (\omega_0)\frac{(\omega-\omega_0)^m}{m!}, \quad
    \beta_m(\omega_0)=\frac{\partial^m \beta}{\partial \omega^m}\Big|_{\omega_0}.
\end{equation}
Retaining only the $m=2$ term, corresponding to group-velocity dispersion (GVD), the dispersion operator becomes $\hat{\mathcal{D}}(\omega)=\beta_2 (\omega-\omega_0)^2/2$, where $\beta_2$ is the GVD parameter.

It is useful to introduce normalized dimensionless variables as $X=x/w_g$, $Y=y/w_g$, $Z=bz$, $T=\tau/T_0$, and $\Psi=E/\sqrt{P_0}$, where $T_0$ has a dimension of time and $P_0$ has a dimension of power.  We choose the center wavelength $\lambda_0 = 1.5~\mu$m at which the GVD coefficient is negative with the value $\beta_2 = -17~\rm ps^2/km$ (see Fig.~\ref{Fig1}(c)). The normalized form  of Eq.~\ref{PropEq1} then becomes
\begin{align} \label{PropEq2}
    i\frac{\partial \Psi}{\partial Z} + \frac{1}{2}\left( \frac{\partial^2 \Psi}{\partial X^2}+\frac{\partial^2 \Psi}{\partial Y^2}\right)+ \frac{\delta_2}{2} \frac{\partial^2 \Psi}{\partial T^2} \nonumber\\  -\frac{1}{2} \left( X^2 + Y^2\right)\Psi   + \gamma |\Psi|^2 \Psi =0 ,
\end{align}
where $\delta_2 = |\beta_2|/bT_0^2$, and $\gamma = \omega_0 n_2 P_0/cbA_{\rm eff}$ are the dimensionless GVD and nonlinear parameters, respectively. One can choose $T_0$ and $P_0$ to make $\delta_2=\gamma=1$. For our GRIN fiber, this choice leads to $T_0=2$~fs and $P_0=4.5$~MW. In the following, we make this choice for numerical analysis but retain these parameters in all equations to identify the effects of dispersive and nonlinear terms.

The presence of radial symmetry in Eq.~(\ref{PropEq2}) in the transverse $X$-$Y$ plane motivates us to consider radially symmetric STOV bullets with the phase singularity at $\rho=0$. In this case, the solution of Eq.~(\ref{PropEq2}) has the form
\begin{equation}\label{gpa3}
    \Psi ( X, Y, Z, T) = \psi(\rho,T) \; \text{exp} (i l\phi +  i\mu Z),
\end{equation}
where $l$ is the topological charge (or OAM) of the STOV bullet, $\mu$ is the propagation constant (or chemical potential in the context of Bose-Einstein condensates), $\psi(\rho,T)$ is a real-valued function governing the spatiotemporal shape of the bullet. Plugging Eq.~(\ref{gpa3}) into Eq. (\ref{PropEq2}), $\psi(\rho,T)$ is found to satisfy
\begin{align} \label{PropEq3}
    \frac{1}{2}\left( \frac{\partial^2 }{\partial \rho^2} + \frac{1}{\rho}
    \frac{\partial }{\partial \rho} - \frac{l^2}{\rho^2}\right) \psi
    + \frac{\delta_2}{2} \frac{\partial^2 \psi }{\partial T^2} \nonumber\\
    -\frac{1}{2} \rho^2 \psi  + \gamma\psi^3  = \mu \psi.
\end{align}
This eigenvalue problem is solved in the next section using a variational technique.

\section{Variational analysis of STOV bullets}

It is worth noting here that Eq.~(\ref{PropEq3}) can be solved by any of the following numerical techniques: (i) a finite-difference scheme, combined with a globally convergent Newton method \cite{Paul2024}, (ii) the split-step Fourier  method used for the NLSE \cite{Wright2015, WrightNature2015,Krupa2017}, or (iii) a modal expansion method resulting in a lrge number of coupled NLSEs \cite{Poletti2008}. All of then are quite time-consuming and require substantial computational resources. Moreover, they hinder physical insight into the formation and evolution of stable STOV bullets. For this reason, we employ the semi-analytical variational analysis that has been widely used for pulse propagation problems in diverse optical systems \cite{Bondeson79,Anderson83, Kaup95,skarka1997,Sahoo17}, including a parabolic GRIN medium \cite{GPA_raghaban_2000,PRivasQB2024,PaulVar23}.

The Lagrangian density associated with Eq.\ (\ref{PropEq3}) is found to be
\begin{align} \label{Ls}
    \mathcal{L}_s = & \mu \rho \big|  \psi \big|^2 + \frac{\rho}{2} \big| \partial_\rho \psi \big|^2 + \frac{\delta_2}{2} \rho \big| \partial_T\psi \big|^2\nonumber\\ &+ \left(  \frac{l^2}{2\rho} + \frac{\rho^3}{2} \right) \big| \psi \big|^2 - \frac{\gamma}{2} \rho \big|  \psi \big|^4.
\end{align}
It is easy to verify that its use in the the Euler--Lagrange equation leads back to Eq.\ (\ref{PropEq3}). Variational analysis requires a suitable \textit{ansatz} for the bullet's shape containing a few parameters that are allowed to change with propagation. To find this ansatz, we note that, without nonlinearity and  dispersion, Eq.\ (\ref{PropEq2}) has exact modal solutions containing Laguerre--Gauss (LG) functions \cite{GPA_grinbook}. In the absence of diffraction and the GRIN potential, Eq.\ (\ref{PropEq2}) supports temporal solitons, with a sech shape for $\gamma > 0$. These insights guide us to use the following ansatz for STOV bullets:
\begin{align} \label{LGpsi}
    \psi(\rho, T) &=  \psi_{0}\,\text{sech}(\eta T) \nonumber\\
    &\times \frac{\rho^{|l|}}{w^{|l|+1}} \exp\left(-\frac{\rho^2}{2 w^2}\right)
     L_p^{|l|} \left( \frac{\rho^2}{w^2} \right),
\end{align}
where $L_p^{|l|}(x)$ is the associated Laguerre polynomial of order $l$ and $p$. The parameters $\psi_0$, $\eta$, and $w$ are real, and their positive values represent the amplitude, inverse temporal width, and spatial width of a STOV bullet, respectively. The ansatz in Eq.\ (\ref{LGpsi}) contains two integers for two degrees of freedom. The azimuthal index $l$ is related to the longitudinal OAM \cite{Allen92} and the radial index $p$ indicates the beam's hyperbolic momentum \cite{PlickPra}. The integer $p$ also determines the number of concentric rings in the LG-mode profile. The case $l=0$ corresponds to a bullet with zero topological charge. This case has been extensively studied in the past \cite{Yu1995,GPA_raghaban_2000, TuritsynPRA18,Ahsan2018,ParraRivasLB23,PRivasQB2024,Sun2024}. In this work, we focus on vortex solutions and assume $l>0$. To find these solutions, the Ritz optimization technique \cite{ParraRivasLB23} is employed. It requires the evaluation of the reduced Lagrangian defined as
\begin{equation}\label{gpa4}
     \bar{L}=\int_{-\infty}^{\infty} dT \int_{0}^{\infty} \mathcal{L}_s(\rho, T) d\rho.
\end{equation}
The integrations over the variables $\rho$ and $T$ are carried out using the ansatz given in Eq.\ (\ref{LGpsi}). Finding a closed-form expression of the reduced Lagrangian has proven difficult for arbitrary values of $p$. For this reason, we focus separately on the cases $p=0$ and $p=1$, for which all integrals can be carried out in a closed form.

\begin{figure*}[tb!]
    \centering\includegraphics[width=\textwidth ]{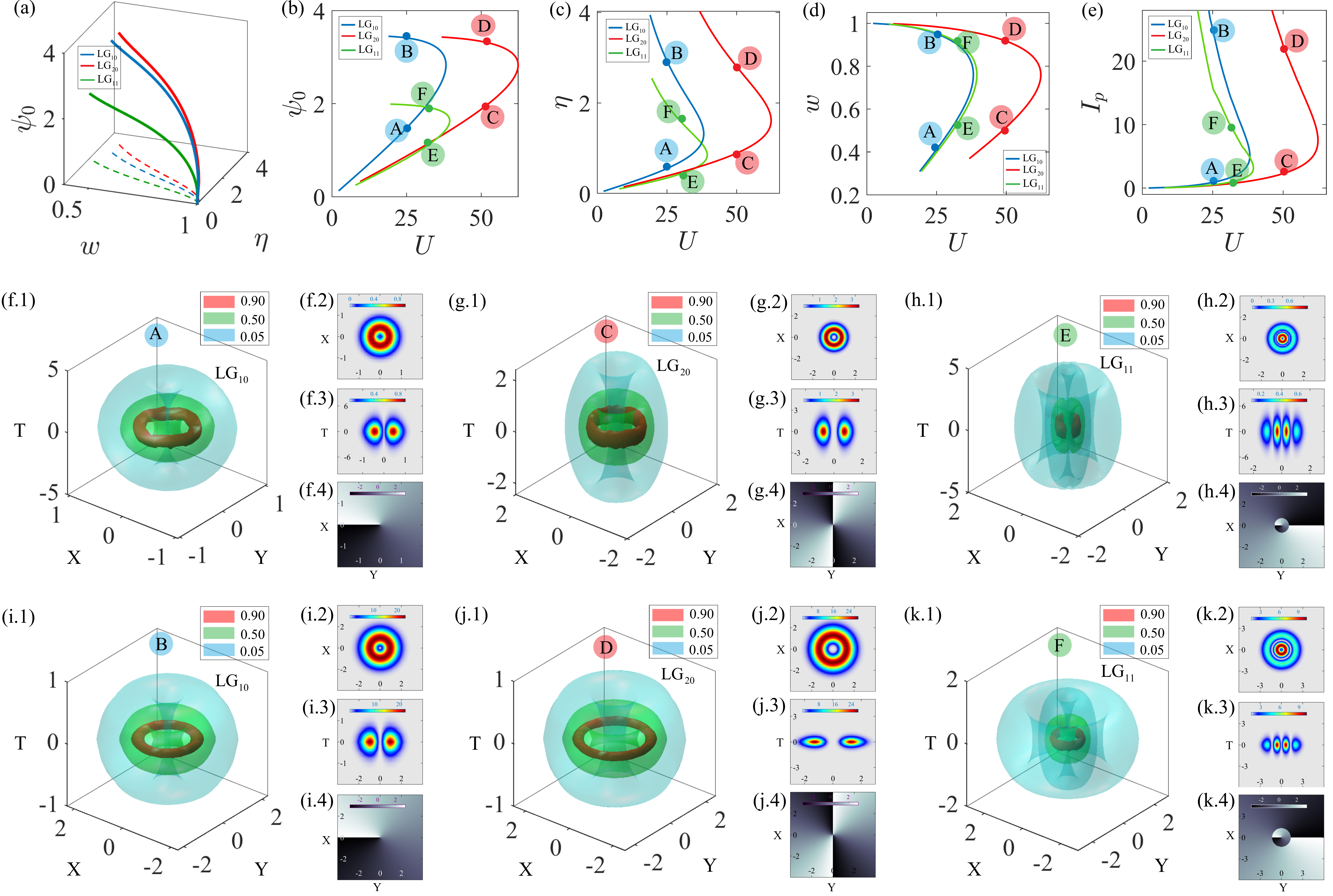}
\caption{Bistable solutions, iso-intensity surfaces, and spatio-temporal intensity distributions for three STOV bullets corresponding to LG$_{10}$ (blue), LG$_{20}$ (red) and LG$_{11}$ (green) modes of a GRIN fiber. {(a)} Amplitude $\psi_0$ (solid lines) and its projection (dashed lines) in the $w$-$\eta$ plane. Bistable solutions depicting {(b)} amplitude $\psi_0$, {(c)} inverse temporal width $\eta$, {(d)} spatial width $w$, and {(e)} peak intensity $I_{\rm p}$ plotted as a function of bullet's energy $U$. Markers A-F in each plot show the values for the six STOV bullets whose shapes and intensities are shown in parts {(f-k)}. Parts (f) and (i) are for the LG$_{10}$ bullets with markers A and B with $U=25$, parts (g) and (j) are for LG$_{20}$ bullets with markers C and D with $U=50$, parts (h) and (k) are for the LG$_{11}$ bullets with markers E and F with $U=30$. In each iso-intensity plot, three intensity levels of $I/I_{\rm p}=0.9$ (red), $0.5$ (green), and $0.05$. Spatial intensity distribution and corresponding phase distribution (in grey) are shown in the plane $T=0$. The spatio-temporal cross-section is depicted in the $X=0$ plane. In all cases, $\delta_2=1$ and $\gamma=1$.}
    \label{Fig2}
\end{figure*}

\subsection{STOV bullets for $p=0$}

The associated Laguerre polynomial for $p=0$ is $L_0^{|l|}(x)=1$ for any integer $l$. Using this value and  and (\ref{LGpsi}), the reduced Lagrangian for $\text{LP}_{l0}$ bullets can be obtained in a closed form after integrating over $\rho$ and $T$, as indicated in Eq.~(\ref{gpa4}). The final result is
\begin{align} \label{Lbarp0}
    \bar L &= \frac{\psi _0^2}{\eta} \left[ \mu  + \frac{\delta_2}{6}\eta^2
    + \frac{(l+1)}{2} \left( w^2+\frac{1}{w^2}\right) \right]\Gamma(l + 1) \nonumber\\
    & -\frac{\gamma}{3}\frac{\Gamma(2l+1)}{2^{2l+1}}\frac{\psi_0^4}{\eta w^2},
\end{align}
where $\Gamma(z)=\int_0^{\infty} e^{-t}t^{z-1} dt$ is the gamma function and $l$ is used to denote $|l|$.

The evolution of three parameters, $\psi_0,\eta$ and $w$, is governed by the reduced Euler--Lagrange equation:
\begin{equation}\label{gpa6}
     \frac{\partial \bar L}{\partial Q}-\frac{d}{dZ}\frac{\partial\bar L}{\partial Q'}=0,
\end{equation}
where $Q$ is one of the parameters and $Q'=dQ/dZ$. The second term becomes zero in the steady state, and the three parameters $\psi_0,\eta$ and $w$ are obtained by setting
\begin{equation}
    \label{ELeffp0}
    \frac{\partial \bar L}{\partial w}=0, \quad \frac{\partial \bar L}{\partial \eta}=0, \quad \frac{\partial \bar L}{\partial \psi_0}=0.
\end{equation}
Exploiting these equations, we obtain three algebraic equations, whose solution provides us with the following three relations for the LG$_{l0}$ bullets:
\begin{subequations} \label{paramsp0} \begin{align}
    w^2 = & \; \frac{\sqrt{4\mu^2+5(l+1)^2}-2\mu}{5(l+1)}, \\
    \eta^2 = & \; {\;\frac{3(l + 1)}{{2{\delta _2}}}\frac{{1 - {w^4}}}{{{w^2}}}}, \\
    \psi_0^2 = & \; \frac{3}{\gamma}\frac{2^{2l}\Gamma(l+2)}{\Gamma(2l+1)}(1-w^4) .
\end{align}\end{subequations}

The preceding expressions elucidate how the GRIN fiber's parameters influence specific characteristics of a STOV bullet. For instance, its amplitude $\psi_0$ depends on the Kerr parameter $\gamma$ but not on the GVD parameter $\delta_2$. In contrast, $\eta$ (related to temporal width) is influenced by the GVD parameter $\delta_2$ only. Moreover, the spatial width $w$ of the STOV depends on neither of these parameters. Figure \ref{Fig2}(a) shows how the STOV's amplitude depends on its spatial and temporal widths for $l=1$ (solid blue) and $l=2$ (solid red), corresponding to the $\text{LG}_{10}$ and $\text{LG}_{20}$ bullets. The bistable nature of these bullets is demonstrated in Figures \ref{Fig2}(b-e), which plot $\psi_0$, $\eta$, $w$, and the peak intensity $I_{\rm p}$) as a function of bullet's energy calculated using
\begin{equation}\label{gpa7}
    U = 2\pi\int_{-\infty}^{\infty} dT \int_0^\infty |\psi(\rho,T)|^2\rho d\rho.
\end{equation}
In all cases, the value of $\mu$ is chosen as a free parameter. A family of STOV bullets exists for a range of $\mu$  the cutoff value of which is determined from the condition $\psi_0^2 > 0$ in Eq. \ref{paramsp0}(c). Note that the energy of $\text{LP}_{l0}$ bullets can be written in a closed form as $U=2\pi(\psi_0^2/\eta)\Gamma(l+1)$.

For a given energy $U$, a STOV bullet can have two different amplitudes and two different widths in space and time. As an example, markers A and B in Figures \ref{Fig2}(b-e) correspond two $\text{LG}_{10}$ bullets having the same energy $U=25$. Similarly, markers C and D represent two $\text{LG}_{20}$ bullets of the same energy $U=50$. Mathematically, this bistability has its origin in multiple roots of the following equation:
\begin{equation}\label{bistab}
    {w^6} - {w^2} + \frac{1}{{{{\left({2\pi} \right)}^2}}}\frac{{{\gamma^2}}}{{3{\delta _2}}}\frac{{{\Gamma ^2}(2l+1)}}{{{2^{4l+1}}\Gamma (l+2){\Gamma^3}(l+1)}}{U^2}=0,
\end{equation}
obtained by substituting the $\eta$ and $\psi_0$ expression  from Eqs.~\ref{paramsp0}(b,c) into $U$. This is a cubic equation in $w^2$ and can be readily solved for a fixed $U$. As an example, for $U=25$ and $l=1$, the two roots of Eq.~\ref{bistab} are $w=0.42$ and $w=0.86$, in agreement with Fig.~\ref{Fig2}(d). The necessary and sufficient condition for finding two distinct real roots of $w$ in the interval $[0,1]$ for a fixed energy $U$ is
\begin{equation}\label{bistabcond}
    0 < {U^2} < \frac{\pi^2}{\sqrt{3}}\frac{\delta _2}{\gamma ^2} \frac{{{2^{4l + 5}}\Gamma (l + 2){\Gamma ^3}(l + 1)}}{{\Gamma ^2}(2l + 1)}.
\end{equation}
The above condition establishes the threshold value of $U$ for the formation of LG$_{l0}$ bullets. For example, for $l=1$ STOV, no bullets are formed with $U >  38$, in agreement with  Figs.~\ref{Fig2}(b-e).

To reveal the intricate nature of the intensity and phase profiles of two LG$_{10}$ bistable STOV bullets, we show their iso-intensity profiles, space-time cross-sections in Figs.~ \ref{Fig2}(f,g,i,j). In each case, the peak intensity $I_{\rm p}$ is normalized to 1, and three iso-surfaces are shown at 90\% (red), 50\% (green), 5\%(cyan) intensity levels. The 3D profiles in  Figs.~ \ref{Fig2}(f.1,g.1,i.1,j.1) reveal doughnut-shape topology of the STOV bullets. The on-axis singularity of each STOV is evident in the spatial intensity distributions (in the $T=0$ plane) shown in Figs.\ \ref{Fig2}(f.2,g.2,i.2,j.2). The spatiotemporal intensity distributions (at $X=0$ plane) are also depicted in Figs.\ \ref{Fig2}(f.3,g.3,i.3,j.3). From the space and time-slice plots, we observe that the LG$_{20}$ bullets have a larger spatial width and lobe separation, compared to the LG$_{10}$ bullets. The spatial phase distributions in the $T=0$ plane are shown in Figs.\ \ref{Fig2}(f.4,g.4,i.4,j.4). As expected, LG$_{10}$ bullets ($l=1$) show a single discontinuity in phase (see parts f.4 and i.4), while LG$_{20}$ bullets ($l=2$) show two discontinuities about the center [see parts g.4 and j.4).

\subsection{STOV bullet for $p=1$}

For $p=1$, the associated Laguerre polynomial becomes $L_1^{|l|}(x)=1+l-x$. Although somewhat more cumbersome, we can still perform the space and time integrations to obtain the following expression for the reduced Lagrangian for $p=1$:
\begin{align} \label{Lbarp1}
    \bar L = & \frac{{\psi _0^2}}{\eta }\left[ {\mu  + \frac{{{\delta _2}}}{6}{\eta ^2} + \frac{{\left( {l + 3} \right)}}{2}\left( {\frac{1}{{{w^2}}} + {w^2}} \right)} \right]\Gamma (l + 2)  \nonumber \\ &  - \frac{\gamma}{3} \frac{{ {2^{ - 2(l + 2)}}(3l + 2)\Gamma (2l + 3)}}{{(2l + 1)}}\frac{{\psi _0^4}}{{{\eta w^2} }}.
\end{align}
Following the same procedure used for $p=0$ case, we obtain the analytic expressions for the three parameters of the LG$_{11}$ STOV bullets:
\begin{subequations} \begin{align} \label{paramsp1}
    w^2 &= \frac{{\sqrt {4{\mu^2} + 5{{\left( {l+3}\right)}^2}} -2\mu}}{{5\left({l+3} \right)}},\\
    \eta^2 &=  \frac{{3\left( {l + 3} \right)}}{{2{\delta _2}}}\frac{{1 - {w^4}}}{{{w^2}}}, \\
    \psi_0^2 &= \frac{3}{\gamma }\frac{{{2^{2l + 3}}(2l + 1)\left( {l + 3} \right)\Gamma (l + 2)}}{{\left( {3l + 2} \right)\Gamma (2l + 3)}}\left( {1 - {w^4}} \right).
\end{align}\end{subequations}
From these expressions, we observe that the bullet parameters have similar dependencies on the GRIN fiber's parameters as in the LP$_{l0}$ case. The total bullet energy in this case is found to be $U=2\pi(\psi_0^2/\eta)\Gamma(l+2)$. The bistability equation for spatial width $w$ in this case can then be derived as
\begin{equation}\label{bistabp1}
    {w^6} - {w^2} + \frac{1}{{{{\left( {2\pi } \right)}^2}}}\frac{{{\gamma ^2}}}{{3{\delta _2}}}\frac{{{{\left( {3l + 2} \right)}^2}{\Gamma ^2}(2l + 3)}}{{{2^{4l + 7}}{{(2l + 1)}^2}\left( {l + 3} \right){\Gamma ^4}(l + 2)}}{U^2} = 0.
\end{equation}
The bistable behavior for the LG$_{l1}$ bullets is depicted by the solid green lines in Figs.\ \ref{Fig2}(a-e). The markers E and F represent two equal-energy $U=30$ LG$_{11}$ bullets. The iso-intensity profiles, space-time slices, and associated phase plots for these two bullets are shown in Figs.\ \ref{Fig2}(h.1-4) and Figs.\ \ref{Fig2}(k.1-4). These plots reveal a multilayered doughnut-shaped topological structure within their spatiotemporal profiles. This is caused by the nonzero radial quantum number $p$ of the Laguerre polynomial. Compared to $p=0$ STOV bullets, nonzero-$p$ bullets are proven to be highly advantageous in photonic computing, especially in quantum information processing. They remain relatively unexplored, and this study may offer a promising approach for their sustenance and stable propagation.

\section{Stability analysis of STOV bullets} \label{VKsec}

\begin{figure*}[tb!]
    \centering\includegraphics[width=\textwidth]{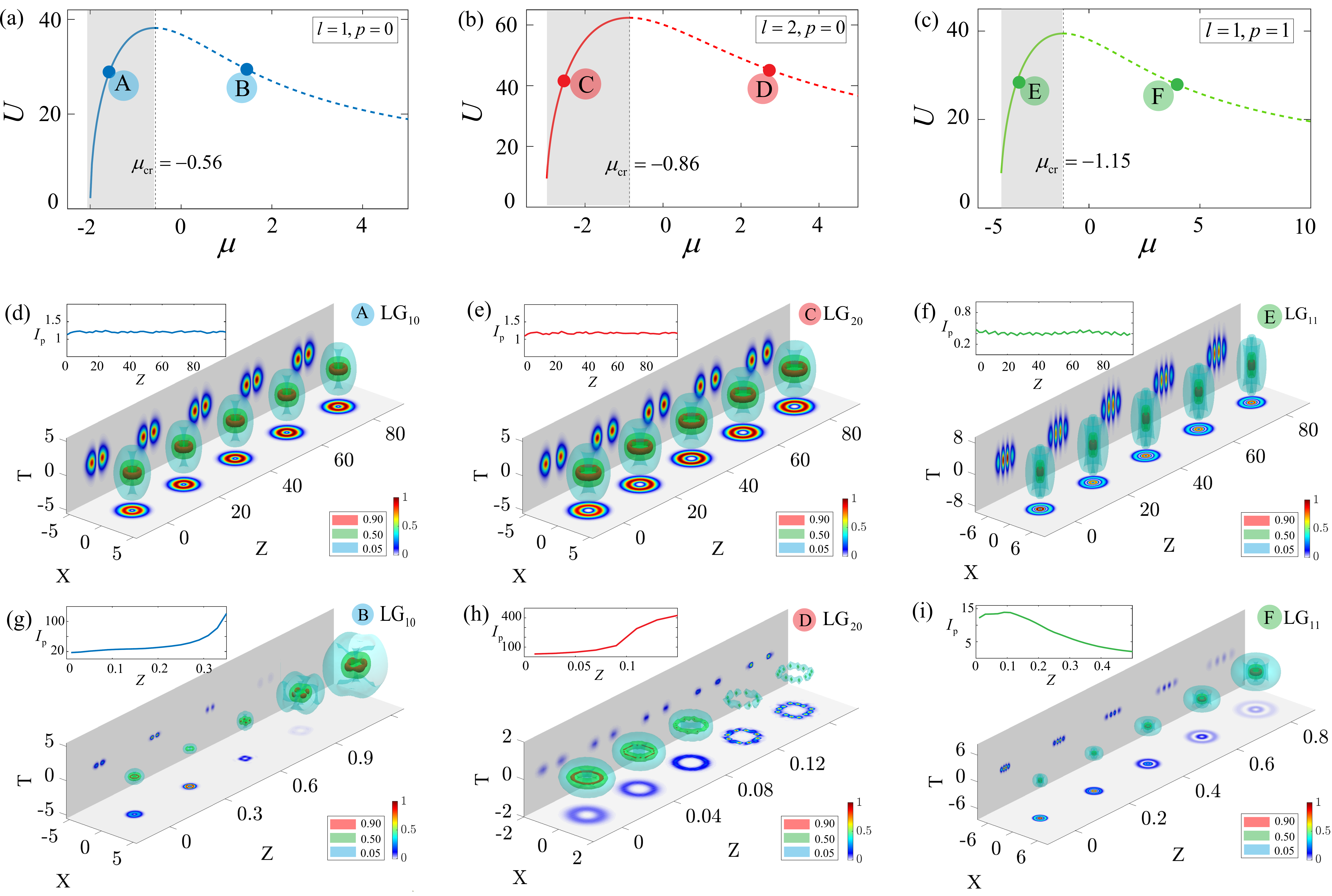}
\caption{Results of VK stability analysis for (left) LG$_{10}$, (center) LG$_{20}$, and (right) LG$_{11}$ bullets. (a-c) Total energy $U$ plotted as a function of $\mu$ for three bullets. Dashed portions indicate instability in each case. Markers A-F indicate the six cases studied numerically, with results shown in parts (d-i). Parts (d-f) show propagation of the three stable bullets (markers A, C, E). Parts (d-f) show the propagation of three unstable bullets (markers B, D, F). In each case, bullet iso-intensity shapes are shown together with their spatial  ($T=0$ plane) and spatiotemporal ($Y=0$ plane) cross-sections. The STOV bullets are launched at $Z=0$ with $5\%$ noise added to their amplitude. Evolution of the peak intensity $I_p$ for each STOV is depicted in the top-left of parts (d-i). Results are obtained with $\delta_2=1$ and $\gamma=1$ }
        \label{Fig3}
\end{figure*}

So far, we have studied the bistable nature and spatiotemporal shapes of the
STOV bullets. In this section, we focus on their stability under small
perturbations. Several methods for soliton stability analysis are available in
the literature, and we choose the Vakhitov-Kolokolov (VK) stability criterion
for our analysis \cite{VKpaper}. It establishes that a soliton state is
linearly stable with respect to small perturbations if the derivative of the
energy $U$ with respect to the propagation constant $\mu$ is a positive
quantity ($dU/d\mu>0$), and unstable otherwise. Plugging the $w$ expression from Eq. \ref{paramsp0}(a) into Eq. \ref{bistab}, it is straightforward to obtain the $\mu$ dependence of $U$ for the $p=0$ bullets:
\begin{equation} \label{Umu} 
U = \sqrt {\frac{1}{{\mathcal{K}\left( l \right)}}\left\{ \begin{array}{l}
\left[ {5{{(l + 1)}^2} - 4{\mu ^2}} \right]\sqrt {4{\mu ^2} + 5{{(l + 1)}^2}} \\
 - \mu \left[ {5{{(l + 1)}^2} - 8{\mu ^2}} \right]
\end{array} \right\}} 
\end{equation}
where the factor $\mathcal{K}(l)$ is defined as
\[\mathcal{K}\left( l \right) = \frac{{{\gamma ^2}}}{{3{\delta _2}}}\frac{{{5^3}{{(l + 1)}^3}}}{{{{\left( {2\pi } \right)}^2}}}\frac{{{\Gamma ^2}(2l + 1)}}{{{2^{4l + 3}}{\Gamma ^3}(l + 1)\Gamma (l + 2)}}.\]
The dependence of energy $U$ on
$\mu$ is computed for the LP$_{l0}$ and LP$_{l1}$ bullets, and the results are depicted
in Fig.\ \ref{Fig3}(a-c).
The solid part of each curve is VK stable
($dU/d\mu>0$), while its dashed part is unstable ($dU/d\mu<0$). 
We observe the
transition from a stable to unstable state occurs at a critical value $\mu_{\rm
cr}$, whose values are $-0.56 , -0.86 $, and $-1.15$ for LG$_{10}$, LG$_{20}$,and LG$_{11}$ bullets, respectively. Although the VK criterion correctly predicts stability under small
perturbations in conservative systems, it is not a sufficient condition for
full stability, especially regarding the splitting instability that arises from
small azimuthal perturbations. To check the robustness of the STOV solutions
bullets, we numerically propagated a perturbed STOV bullet over long distances
using the SSFM algorithm, after adding $5\%$ random noise to its input profile,
and the results are shown in Fig.~\ref{Fig3}, where we show both the
iso-intensity and space-time projection plots in six different situations.
Stable propagation of LG$_{10}$, LG$_{20}$, and LG$_{11}$ bullets is shown in
parts (d), (e) and (f), respectively, and the results are consistent with the
VK predictions. The propagation continues to remain stable beyond $Z=60$, which
corresponds to 375 diffraction lengths using $L_{\rm DF}= b\beta_0 w_0^2 =
0.16$ . For completeness, VK-predicted
unstable propagation also shown in parts (g), (h) and (i), respectively. Note
how the bullets are split into several fragments before collapsing within one
diffraction length.

\section{Breathing dynamics of STOV Bullets}

The study of STOV bullets in the preceding sections has provided valuable insights. However, exploring their dynamics beyond the steady state is vital for assessing the robustness of bullets under a wide range of perturbations. In this section, we study the breathing dynamics of STOV bullets by incorporating the $Z$-dependence into the static Lagrangian given in Eq.~\ref{Ls}. More specifically, we modify this Lagrangian as
\begin{align} \label{Ld}
    \mathcal{L}_D = & \frac{i}{2}\rho \left( {\psi {\partial _Z}{\psi ^*} - {\psi ^*}{\partial _Z}\psi } \right) + \frac{\rho }{2}{\left| {{\partial _\rho }\psi } \right|^2} + \frac{{{\delta _2}}}{2}\rho {\left| {{\partial _T}\psi } \right|^2}   \nonumber \\ &   + \left( {\frac{{{l^2}}}{{2\rho }} + \frac{{{\rho ^3}}}{2}} \right){\left| \psi  \right|^2} - \frac{\gamma }{2}\rho {\left| \psi  \right|^4}.
\end{align}
It is easy to show that Eq.~(\ref{PropEq2}) is recovered when this Lagrangian is used in the Euler--Lagrange equation, with $\psi^*$ acting as the variable for differentiation.

For the variational analysis, we need to modify the ansatz used earlier in Eq.~(\ref{LGpsi}). The reason is that curvature of the phase front has to be included in both the space and time domains. Thus, we choose the following \textit{ansatz} with six parameters, all of which are allowed to vary with $Z$:
\begin{align} \label{LGdynamic}
    \psi &= \psi_{0}(Z) \,\text{sech}(\eta(Z) T) \frac{\rho^{|l|}}{w(Z)^{|l|+1}} L_p^{|l|} \left( \frac{\rho^2}{w(Z)^2} \right)  \nonumber \\  & \times \exp\left(-\frac{\rho^2}{2 w(Z)^2} + iC(Z) T^2 + i D(Z)\rho^2 + i \Theta(Z) \right).
\end{align}
Here, $C(Z)$ and $D(Z)$ are the temporal and radial chirp parameters and $\Theta(Z)$ is the phase of STOV bullets.

As before, the reduced Lagrangian can is calculated by carrying out the integrations indicated in Eq.~(\ref{gpa4}). For $p=0$ corresponding to LP$_{l0}$ bullets, the result is found to be
\begin{align} \label{LDdynamicp0}
    &{\bar L_D} = \frac{{\psi _0^2(Z)}}{\eta(Z) }\left[ {\frac{{{\pi ^2}}}{{12}} \frac{C'(Z)}{{{\eta^2(Z)}}} + (l + 1){w^2(Z)}D'(Z) + \Theta'(Z)} \right] \nonumber\\
    & \times \Gamma (l + 1) + \frac{{{\delta _2}}}{6}\frac{{\psi_0^2(Z)}}{\eta(Z) }\left[ {\frac{{{\pi ^2}}}{{{\eta^2(Z)}}}{C^2(Z)} + {\eta^2(Z)}} \right]\Gamma (l + 1) \nonumber\\
    & + \frac{{\psi_0^2(Z)}}{{2\eta(Z) }}\left[ {4{D^2(Z)}{w^2(Z)} + {w^2(Z)} + \frac{1}{{{w^2(Z)}}}} \right]\Gamma (l + 2) \nonumber\\ & - \frac{\gamma }{3}
    \frac{ \Gamma (2l+1)}{2^{ 2l+1}}\frac{{\psi_0^4(Z)}}{{{w^2(Z)}\eta(Z) }},
\end{align}
where the symbol $'$ stands for a derivative with respect to $Z$, i.e., $Q'=dQ/dZ$.

Next, we use the Euler-Lagrange equation in Eq.~(\ref{gpa6}) for $Q=\psi_0, \eta, w, C, D$ and $\Theta$, and obtain the following six ordinary differential equations describing the evolution of six parameters:
\begin{subequations}\label{Evolutionp0} \begin{align}
    \frac{{d{\psi _0}}}{{dZ}} = & \;  - {\delta _2}C{\psi _0} , \\
    \frac{{d\eta }}{{dZ}}  = & \;  - 2{\delta _2}C\eta ,  \\
    \frac{{dw}}{{dZ}} = & \; 2Dw , \\
    \frac{{dC}}{{dZ}} = & \;  - 2{\delta _2}\left( {{C^2} - \frac{{{\eta ^4}}}{{{\pi ^2}}}} \right) - \frac{\gamma }{\pi ^2} \frac{{{2^{ - 2l }}\Gamma (2l+1)}}{{ \Gamma (l+1)}}\frac{{{\eta ^2}\psi _0^2}}{{{w^2}}} , \\
    \frac{{dD}}{{dZ}} = & \;  - 2{D^2} - \frac{1}{2} \left( 1 - \frac{1}{w^4} \right)  -  \frac{\gamma}{3}\frac{ {{2^{ - 2l-1 }}\Gamma (2l+1)}}{{\Gamma (l+2)}}\frac{{\psi _0^2}}{{{w^4}}}, \\
    \frac{{d\Theta}}{{dZ}} = & \; -\frac{1}{3}\delta_2 \eta^2 - \frac{(l+1)}{w^2} + \frac{7\gamma}{3} \frac{2^{-2l-2} \Gamma(2l+1)}{\Gamma(l+1)} \frac{\psi_0^2}{w^2}
\end{align}\end{subequations}
As the phase $\Theta$ does not affect other parameters, we can ignore the last equation, reducing the 6D system to a 5D one. We emphasize that a numerical solution of five coupled differential equations is much less time-consuming and provides physical insight, compared to full numerical simulations based on Eq.\ (\ref{PropEq2}). For example, it is easy to show from Eqs.\ \ref{Evolutionp0}(a,b) that
\begin{equation}
    \label{Econserve}
    \frac{d}{dZ} \left( \frac{\psi_0^2}{\eta} \right)=0.
\end{equation}
Noting the total energy is $U$ scales as $\psi_0^2/\eta$, Eq.\ (\ref{Econserve}) implies the conservation of energy for LG$_{l0}$ STOVs.

The steady-state STOV parameters appearing in Eqs.\ (\ref{paramsp0}) can be retrieved by setting all $Z$ derivatives to zero in Eqs.\ (\ref{Evolutionp0}) and using $\mu$ for $\Theta'$. Let us denote this steady-state point by the column vector $q_e=\left(\psi_{e}, \eta_{e}, w_{e}, C_e, D_e\right)^{\rm T}$ in the 5D parameter space, where the suffix $e$ stands for equilibrium. Note that $C_e=0$ and $D_e=0$ at this point. To study the dynamics at nearby points not too far from the equilibrium, we consider the solution at the point $q=q_e + \delta q$ where
$\delta q =\left( \Delta \psi_0, \Delta \eta, \Delta w, \Delta C, \Delta D\right)^{\rm T}$ is a perturbation vector in the parameter space satisfying $|\delta q|/q_e\ll 1$. The dynamics of the system are captured by linearizing Eqs.\ (\ref{Evolutionp0}) in terms of $\delta q$. This leads to the following matrix equation:
\begin{equation}\label{dynL}
    \frac{d\delta q}{dz} = \mathcal{J}[q_e]\delta q,
\end{equation}
where $\mathcal{J}[q_e]$ is the $5 \times 5$ Jacobian matrix such that
\begin{equation}\label{Jacob}
    \mathcal{J}_{ij} = \left( \frac{\partial f_i}{\partial q_j}\right) \Big|_{q_e}.
\end{equation}
Here, $f=(f_1,f_2,f_3,f_4,f_5)^{\rm T}$ is a column vector consisting of functions on the right side of Eqs.\ \ref{Evolutionp0}. For example, $f_1=- {\delta_2}C{\psi _0}$, $f_2=- 2{\delta _2}C\eta$, and so on. The elements of the Jacobian matrix in Eq.\ (\ref{Jacob}) are evaluated at the equilibrium point $q_e$ (see Appendix~\ref{appendA} for their expressions).

Stability of the dynamical system governed by Eqs.\ (\ref{Evolutionp0}) can thus be evaluated by solving the eigenvalue problem
\begin{equation}
    \label{eigs}
    \mathcal{J} v = \Lambda v,
\end{equation}
where $\Lambda$ and $v$ are, respectively, the eigenvalues and eigenvectors associated with the Jacobian matrix $\mathcal{J}$. The eigenvalues satisfy the characteristic fifth-order polynomial equation of the form $\sum_{k=1}^5c_k\Lambda^k=0$ where the coefficients $c_k$'s are functions of the equilibrium bullet parameters $\psi_{0e}, \eta_e, w_e$. The five roots, $\Lambda_{i=1,..,5}$ of this polynomial provide five eigenvalues that can be complex. Stability is governed by their real parts. If the real part is negative for all eigenvalues, the system is stable. The maximum value of real parts, if positive, indicates instability.

\begin{figure}[tb!]
    \centering\includegraphics[width=0.5\textwidth]{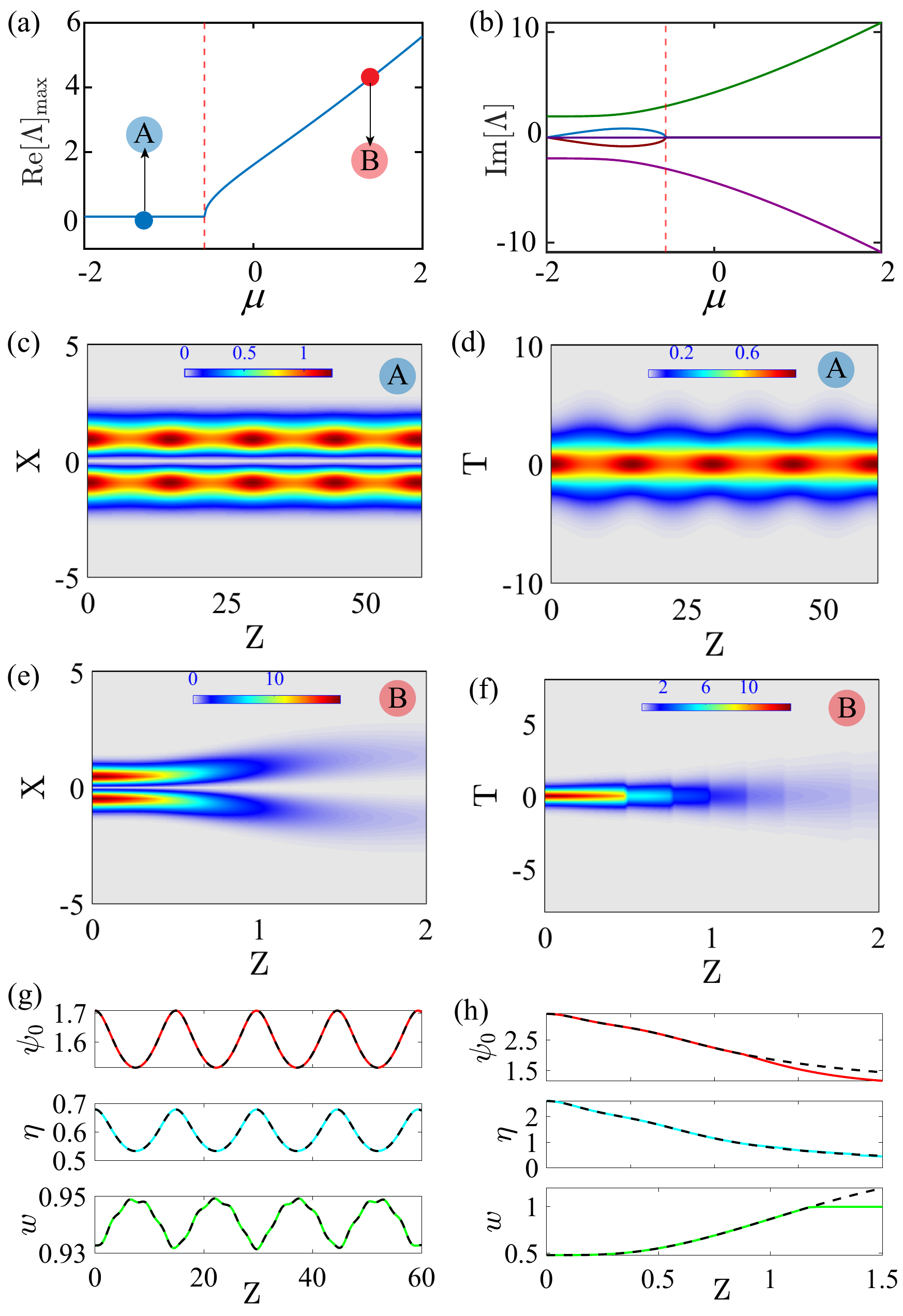}
\caption{\textbf{(a)} $\rm Re[\Lambda]_{max}$ and \textbf{(b)} imaginary parts $\rm Im[\Lambda]$ of all eigenvalues for the LG$_{10}$ STOV bullets. The vertical dashed line marks the critical value $\mu_{\rm cr}=-0.56$ above which bullets become unstable. Markers A and B show the two cases studied numerically in parts (b-h). At point A, stable oscillations both in (c) space and (d) time. At point B, the unstable bullet starts spreading in both (e) space and (f) time. Parts (g) and (h) compare numerical results (solid line) with variational predictions (dashed lines) in the cases A and B, respectively, by plotting $\psi_0$, $\eta$, and $w$ as a function of $z$.
Results are obtained using $\delta_2=1$ and $\gamma=1$.} \label{Fig4}
\end{figure}

Figure \ref{Fig4}(a) shows max(Re[$\Lambda$]) as a function of $\mu$ for LG$_{10}$ STOV bullets, and part (b) shows imaginary parts of all five eigenvalues. The maximum value in part (a) is zero for $\mu<-0.56$. This correspond to neutral stability as perturbations neither decay nor grow with propagation, but exhibit oscillations around the steady-state at a frequency set by the imaginary part of an eigenvalue. We emphasize that the value $\mu_{\rm cr}=-0.56$ is in agreement with the VK prediction in Section \ref{VKsec}. In the terminology of nonlinear dynamics, the fixed points are referred to as ``centers," around which oscillations occur. Such oscillations of the parameters are referred to as breathing of a STOV bullet. In the unstable region for $\mu>\mu_{\rm cr}$, eigenvalues exhibit a positive real part, leading to an exponential growth of perturbations.

To verify the predictions of preceding linear stability analysis, we used the SSFM and solved numerically the underlying governing propagation (\ref{PropEq3}) for two cases marked as A and B in Figure \ref{Fig4}(a), and the results are shown in Figs.~\ref{Fig4}(c-f). We also solved the coupled ODEs (\ref{Evolutionp0}) with the fourth-order Runge-Kutta (RK4), and the results are shown in parts (g) and (h) of the same figure. Specifically, the stable LG$_{10}$ bullet at point A exhibits stable breathing with periodic oscillations, both in space and time, as seen in parts (c) and (d) of Fig.~\ref{Fig4}. In contrast, as seen in parts (e) and (f), at the unstable point B, the bullet starts spreading in both space and time. See the Supplemental Movies \cite{SMV3} and \cite{SMV4} for iso-intensity evolution of these two bullets. The dashed lines in parts (g) and (h) compare the variational predictions with full simulation results shown by solid lines.

Variational analysis can be used to provide deeper insight into the stable breathing of the bullet parameters. We used the evolution equations (\ref{Evolutionp0}) to obtain the following second-order differential equation \cite{skarka1997} for the spatial width $w$:
\begin{equation}\label{Wevo}
    \frac{d^2w}{dZ^2} + w = \frac{1}{w^3} - \frac{\text{K}_0\gamma}{3} \frac{2^{-2l} \Gamma(2l+1)}{\Gamma(l+2)} \frac{\eta}{w^3},
\end{equation}
where $\text{K}_0=\psi_0^2/\eta$ is a constant of motion related to the bullet's energy from Eq.(\ref{Econserve}). Assuming negligible variations in the temporal width, we set $\eta = \eta_e$. In this case, Eq.~\ref{Wevo} can be solved to obtain the following closed-form solution for the spatial width of the STOV bullet:
\begin{equation}\label{Wevosol}
    w(z)=\sqrt{w_0^2 \cos^2z + \frac{\alpha}{w_0^2} \sin^2 z},
\end{equation}
where $w_0$ is the initial width and $\alpha$ is defined as
\begin{equation}\label{Wevosol}
    \alpha=1-(\text{K}_0/3)\gamma 2^{-2l} [\Gamma(2l+1)/\Gamma(l+2)]\eta_e.
\end{equation}

Further insight can be gained by integrating Eq.~(\ref{Wevo}) once to obtain an equation similar to that of a particle trapped in a potential well:
\begin{equation}
    \label{WevoPot}
   \frac{1}{2}\left( \frac{dw}{dZ} \right)^2 + \Omega_w(w) =  \Omega_w(w_0),
\end{equation}
where the potential function $\Omega_w(w)$ is given by
\begin{equation}
    \label{WPot}
   \Omega_w(w) = \frac{w^2}{2} + \frac{1}{2w^2}\left[1-\frac{  \gamma \text{K}_0 \eta_e 2^{-2l}\Gamma(2l+1)}{3\Gamma(l+2)} \right] .
\end{equation}
Here $w_0$ denotes the initial spatial width of the STOV with a planar wave front ($D=0$).

\begin{figure}[t!]
    \centering\includegraphics[width=0.5\textwidth]{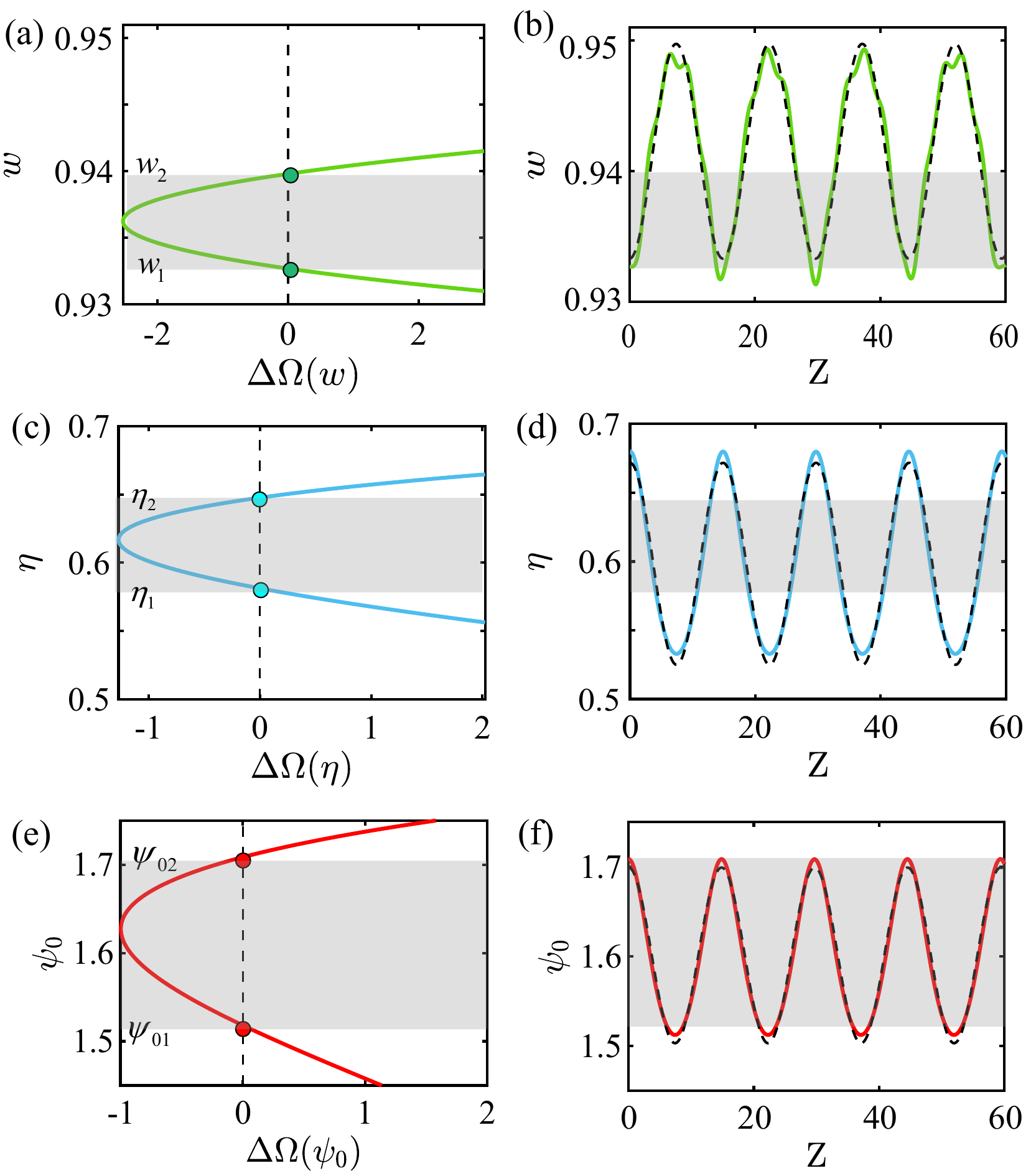}
\caption{Potentials (left) and periodic evolution (right) of the three parameters of a stable LG$_{10}$ STOV bullet, launched with an initial perturbation $\Delta\eta(0)=0.05 \eta_e$ (no perturbation for other parameters). Dashed lines in parts (b,d,f) show the analytical approximation for comparison. The shaded region in each plot marks the upper and lower bounds, as predicted by the turning points of the corresponding potential. Results are obtained for $\delta_2=1$ and $\gamma=1$.} \label{Fig5}
\end{figure}

By comparing Eq.~(\ref{WevoPot}) to that governing the motion of a particle in a potential field, we observe that  $\Omega_w(w)$ plays the role of an effective potential for the radial width $w$. Proceeding with the analogy of a particle trapped in a potential well, we gain a deeper understanding of the nonequilibrium STOV dynamics. For example, under sustainable stable oscillations, the radial width $w$ lies between two boundaries $w_1$ and $w_2$ such that $w_1<w<w_2$, where the turning points $w_1$ and $w_2$ are positive real roots of the equation $\Omega_w(w)=\Omega_w(w_0)$. These turning points depend on the initial conditions and the waveguide parameters (dispersion, nonlinearity). The equilibrium width $w_e$ is obtained from $d\Omega_w/dZ \lvert_{w=w_e}=0$.

Figure \ref{Fig5}(a)  depicts the potential $\Omega_w$ and the corresponding turning points for a stable LG$_{10}$ STOV with $\mu=-1.6$. We find $w_1=0.93 $ and $w_2=0.94$ from the zero-crossing points of the potential difference $\Delta \Omega(w)=\Omega(w)-\Omega(w_0)$. Comparing with the VA evolution depicted in part (b), we see that the lower bound agrees closely with the VA-predicted bound $w_1^{\rm (VA)}=0.933$, while the upper bound $w_2^{\rm (VA)}=0.95$ deviates by $\sim 1\%$. Similarly assuming $w \simeq w_e$ throughout propagation, we further derive the approximate potential functions  $\Omega(\eta)$ and $\Omega(\psi_0)$ associated with the parameters $\eta$ and $\psi_0$:
 \begin{subequations} \label{etapsiPot} \begin{align}
    \Omega_\eta(\eta) = & \;\frac{{2\delta _2^2}}{{3{\pi ^2}}}{\eta ^6} - \frac{{\gamma {{\rm K}_0}{\delta _2}}}{{5{\pi ^2}}}\frac{{{2^{ - 2l + 1}}\Gamma (2l+1)}}{{w_e^2\Gamma (l+1)}}{\eta ^5}  , \\
    \Omega_\psi(\psi_0) = & \; {\frac{{\delta _2^2}}{{5{\pi ^2}{\rm K}_0^4}}\psi _0^{10} - \frac{{\gamma {\delta _2}}}{{{\pi ^2}{\rm K}_0^2}}\frac{{{2^{ - 2l - 3}}\Gamma (2l+1)}}{{w_e^2\Gamma (l+1)}}\psi _0^8}.
\end{align}\end{subequations}

Part (c) of Figure \ref{Fig5} illustrates the $\eta$-potential difference $\Delta \Omega(\eta)=\Omega(\eta)-\Omega(\eta_0)$ and the associated turning points at $\eta_1=0.582$ and $\eta_2=0.648$. The $\eta(z)$ evolution obtained from the solution of Eq.\ \ref{Evolutionp0}(b) is shown in part (d) by a solid line, with the lower and upper oscillation limits of $\eta_1^{\rm (VA)}=0.525 $ and $\eta_2^{\rm (VA)}=0.68$, respectively. Similarly, the $\psi_0$ potential difference $\Delta \Omega(\psi)$ with turning points $\psi_{01}=1.5$, $\psi_{02}=1.7$ and the corresponding RK4 solution $\psi_0(Z)$ obtained from Eq.~\ref{Evolutionp0}(a) are shown in parts (e) and (f), respectively. Note that the potential formalism works better for predicting the bounds on $\psi_0$, compared to $w$ and $\eta$. This is because the coefficient of the dominant term ($\psi_0^{10}$) in the potential is unaffected by the assumption $w \simeq w_e$.

The breathing frequency of the STOV can be assessed by studying small-amplitude disturbances around the equilibrium, i.e., $ w = w_e + \Delta w$. Assuming, $\Delta w / w_e\ll 1$, and linearizing Eq.~(\ref{Wevo}) in  $\Delta w$ leads to a simple harmonic oscillator equation in $\Delta w$
\begin{equation}
    \label{delWevo}
    \frac{d^2(\Delta w)}{dZ^2} + \omega_w^2 (\Delta w) = 0,
\end{equation}
which has the analytical solution $\Delta w(Z)= \Delta w(0)\cos(\omega_w Z)$, where the oscillation frequency $\omega_w$ is given by
\begin{equation}
    \label{bp}
   \omega_w = \sqrt{\frac{d^2\Omega_w}{dw^2}\Biggl|_{w_e}}
\end{equation}
For stable oscillation, $\omega_w^2$ has to be positive. Exploiting the functional form of $\Omega_w$ in Eq.~(\ref{WPot}), one can show that the breathing frequency is given by,
\begin{equation}
    \label{wperiod}
  \omega_w=\sqrt{1+ \frac{3}{w_e^4} -\frac{\gamma \text{K}_0\eta_e   2^{-2l} \Gamma (2 l+1)}{\Gamma (l+2)w_e^4}  }.
\end{equation}

Inserting the numerical values, we obtain $\omega_w=2$, a value that deviates significantly from the  VA-predicted value of $\omega_w^{\rm (VA)}=0.42$, shown by a solid line in Fig.\ \ref{Fig5}(b). This large mismatch in frequency arises from the approximation that the temporal width remains close to the equilibrium value, i.e., $\eta = \eta_e$. However, this is not the case, as shown in Fig.~\ref{Fig5}(d), where $\eta$ oscillates at a frequency close to that of the $w$ oscillations.

To obtain a more accurate estimation of oscillation frequency, we adopt a matrix-diagonalization technique that takes into account the evolution in all three parameters ($\Delta \eta,~\Delta w$ and $\Delta \psi_0$). Taking the derivative of Eq.~\ref{dynL}) with respect to $Z$, we obtain the following second-order differential equation:
  \begin{equation}
    \label{dynL2}
   \left( \frac{d^2 }{dZ^2} + \mathcal{A} \right) \delta q=0,
\end{equation}
where the coefficient matrix $\mathcal{A}=-\mathcal{J}^2[q_e]$ is non-diagonal. To make $\mathcal{A}$ diagonal, we adopt the eigenvalue decomposition method $\mathcal{A}_D= \mathcal{P}_e^{-1} \mathcal{A} \mathcal{P}_e$, where ${\mathcal{P}_e}$ is a matrix whose columns are the eigenvectors of $\mathcal{A}$ and $\mathcal{A}_D$ is the corresponding diagonal matrix (see Appendix \ref{appendB} for derivation). Focusing only on the $\eta, w$ and $\psi_0$ evolution, ${\mathcal{P}_e}$ reduces to a $3 \times 3$ matrix. The solution to the reduced problem of three independent simple harmonic oscillators is then readily obtained:
\begin{equation}
   \label{DelwetapsievoAnalytical}
    \left( {\begin{array}{*{20}{c}}
    {\Delta \eta}\\ {\Delta w}\\ {{\Delta \psi _0}}
    \end{array}} \right) =   {\mathcal{P}_e}\left( {\begin{array}{*{20}{c}}
    {\Delta \tilde \eta (0)\cos \left( {{\omega _1}Z} \right) + \frac{{{F_1}}}{{\omega _1^2}}\left( {1 - \cos \left( {{\omega _1}Z} \right)} \right)}\\
    {\Delta \tilde w(0)\cos \left( {{\omega _2}Z} \right) + \frac{{{F_2}}}{{\omega _2^2}}\left( {1 - \cos \left( {{\omega _2}Z} \right)} \right)}\\
    {\Delta {{\tilde \psi }_0}(0)\cos \left( {{\omega _3}Z} \right) + \frac{{{F_3}}}{{\omega _3^2}}\left( {1 - \cos \left( {{\omega _3}Z} \right)} \right)}
    \end{array}} \right) ,
\end{equation}
where the tilde variables are related to the untilde ones as $[\Delta \tilde{\eta}, \Delta \tilde{w}, \Delta \tilde{\psi}_0]^{\rm T}=P_e^{-1} [\Delta {\eta}, \Delta {w}, \Delta {\psi}_0]^{\rm T}$. The frequencies $\omega_i$ and amplitudes $F_i$ $(i=1,2,3)$ depend only on the equilibrium parameters (see Appendix \ref{appendB} for their  expressions). We emphasize that the above expressions are obtained with the assumption of zero initial radial and temporal chirps. In Figs.~\ref{Fig5}(b,d,f), we show the predictions of this analysis by dashed lines. Good agreement seen there justifies our approach adopted here.

\section{Conclusions}

In this article, we explore theoretically the formation, stability, and
evolution of STOV bullets, characterized by distinct azimuthal and radial
quantum numbers, inside a GRIN multimode fiber. Exploiting a rigorous
variational treatment that is also supported by full numerical simulations of the underlying (3+1)D NLSE, we predict the formation of bistable STOV bullets under experimentally accessible conditions. We focus our investigation on how the shape of such vortex bullets varies with control parameters and discuss how their spatial and temporal structure evolves with changes in control parameters.

The stability of the vortex bullets is analyzed through an analytical technique
based on the VK criteria and a numerical technique based on small perturbations
of the steady-state parameters of a vortex bullet. We have identified a cut-off
value of the propagation constant that delineates stability; bullets remain
stable below this cut-off value but become unstable for values above it. To
gain a deeper understanding of stability dynamics, we perturb the stationary
solution from the steady state by changing its five parameters and study their
evolution inside the GRIN fiber. Analyzing the full eigenspectrum of the
perturbed bullets, we demonstrate that STOV bullets become unstable beyond a
critical value of the propagation constant, which is found to agree with the VK
prediction. In all cases, we validate the variational results through full
numerical simulations of the (3+1)D NLSE\@.

To develop intuition about the non-stationary evolution of vortex bullets, we
develop a potential formalism for the pulse parameters that is analogous to a
point particle trapped in a potential. We show that the perturbation amplitudes
exhibit stable oscillations about the equilibrium point under certain
conditions set by parameters of the GRIN fiber. This formalism allows us to
estimate the oscillation frequency and the range over which the amplitude varies.
Using a matrix-diagonalization approach, we further derive a closed-form
expressions of the bullet parameters that more accurately track the evolution
under perturbation. These results enhance our understanding of self-trapped,
spatiotemporal vortex solitons in a graded-index medium and may lead to
potential applications in the area of classical and quantum information
processing.

\section*{Acknowledgement}
A.P. would like to thank IIT Kharagpur and University of L'Aquila for providing funding and computational resources to carry out this work.

\section*{Data Availability}
The data that support the findings of this article are not publicly available. The data are available from the authors upon reasonable request.

\appendix
\section{Jacobian matrix expression in Eq.~\ref{Jacob}} \label{appendA}

The full form the Jacobian matrix introduced in Section V is
\begin{equation} \label{JacobExpress}
\mathcal{J}[q_e] = \left[ {\begin{array}{*{20}{c}}
0&0&0&{ - {\delta _2}{\psi_{e}}}&0\\
0&0&0&{ - 2{\delta _2}{\eta_e}}&0\\
0&0&0&0&{2{w_e}}\\
{{\mathcal{F}_{41}}}&{{\mathcal{F}_{42}}}&{{\mathcal{F}_{43}}}&0&0\\
{{\mathcal{F}_{51}}}&0&{{\mathcal{F}_{53}}}&0&0
\end{array}} \right]
\end{equation}
where $\mathcal{F}_{ij}$ are functions of the steady-state parameters and are
given by:
\begin{subequations} \label{Fijp0}
\begin{align*}
\mathcal{F}_{41} = & \;  - \frac{\gamma }{{{\pi ^2}}}\frac{{{2^{ - 2l + 1}}\Gamma (2l+1)}}{{\Gamma (l+1)}}\frac{{\eta _e^2{\psi _{e}}}}{{w_e^2}}, \\
\mathcal{F}_{42} = & \; \frac{8{{\delta _2}}}{{{\pi ^2}}}{\eta_e}^3 - \frac{\gamma }{{{\pi ^2}}}\frac{{{2^{ - 2l + 1}}\Gamma (2l+1)}}{{\Gamma (l+1)}}\frac{{{\eta _e}\psi _{e}^2}}{{{w_e}^2}}, \\
\mathcal{F}_{43} = & \; \frac{\gamma }{{{\pi ^2}}}\frac{{{2^{ - 2l + 1}}\Gamma (2l+1)}}{{\Gamma (l+1)}}\frac{{{\eta _e}^2\psi _{e}^2}}{{w_e^3}}, \\
\mathcal{F}_{51} = & \;  - \frac{\gamma}{3} \frac{{{2^{ - 2l }}\Gamma (2l+1)}}{{\Gamma (l+2)}}\frac{{{\psi _{e}}}}{{w_e^4}}, \\
\mathcal{F}_{53} = & \;  - \frac{2}{{w_e^5}} + \frac{\gamma}{3} \frac{{{2^{ - 2l + 1}}\Gamma (2l+1)}}{{\Gamma (l+2)}}\frac{{\psi _{e}^2}}{{w_e^5}}.
\end{align*}\end{subequations}

\section{Derivation of analytical expressions in Eq.~\ref{DelwetapsievoAnalytical}} \label{appendB}

Taking the second derivatives of Eqs.~\ref{Evolutionp0}(a,b,c) and simplifying,
we obtain the following equations:
 \begin{subequations} \label{Potetapsi} \begin{align}
\frac{{{d^2}\eta }}{{d{Z^2}}} = & \; 8\delta _2^2{C^2}\eta  - \frac{4}{{{\pi ^2}}}\delta _2^2{\eta ^5} + \frac{{\gamma {\delta _2}{{\rm K}_0}}}{{{\pi ^2}}}\frac{{{2^{ - 2l + 1}}\Gamma (2l+1)}}{{\Gamma (l+1)}}\frac{{{\eta ^4}}}{{{w^2}}} , \\
\frac{{{d^2}w}}{{d{Z^2}}} = & \;  - w + \frac{1}{{{w^3}}} - \frac{\gamma }{3}\frac{{{2^{ - 2l}}\Gamma (2l + 1)}}{{\Gamma (l + 2)}}\frac{{\psi _0^2}}{{{w^3}}} , \\
\frac{{{d^2}\psi_0}}{{d{Z^2}}} = & \; 3\delta _2^2{C^2}{\psi _0} - \frac{{2\delta _2^2}}{{{\pi ^2}{\rm K}_0^4}}\psi _0^9 + \frac{{\gamma {\delta _2}}}{{{\pi ^2}{\rm K}_0^2}}\frac{{{2^{ - 2l }}\Gamma (2l+1)}}{{\Gamma (l+1)}}\frac{{\psi _0^7}}{{{w^2}}}&
\end{align}\end{subequations}
Using a Taylor expansion of the three parameters around their equilibrium
values $\eta_e, w_e, \psi_e$ with $C_e=0$, we obtain the following matrix
equation:
\begin{equation} \label{coupledetaw}
  \frac{{{d^2}}}{{d{Z^2}}}\left[ {\begin{array}{*{20}{c}}
{\Delta \eta }\\
{\Delta w}\\
{\Delta {\psi _0}}
\end{array}} \right] = \left[ {\begin{array}{*{20}{c}}
{{f_{1e}}}\\
{{f_{2e}}}\\
{{f_{3e}}}
\end{array}} \right] + \left[ {\begin{array}{*{20}{c}}
{{A_{11}}}&{{A_{12}}}&0\\
{{A_{21}}}&{{A_{22}}}&A_{23}\\
0&{{A_{32}}}&{{A_{33}}}
\end{array}} \right]\left[ {\begin{array}{*{20}{c}}
{\Delta \eta }\\
{\Delta w}\\
{\Delta {\psi _0}}
\end{array}} \right]
\end{equation}
where the nonzero matrix elements are
\[\begin{array}{l}
{f_{1e}} =  - \frac{4}{{{\pi ^2}}}\delta _2^2\eta _e^5 + \frac{{\gamma {\delta _2}{{\rm K}_0}}}{{{\pi ^2}}}\frac{{{2^{ - 2l + 1}}\Gamma (2l+1)}}{{\Gamma (l+1)}}\frac{\eta_e^4}{w_e^2},\\ \\
{f_{2e}} =  - {w_e} + \frac{1}{{w_e^3}} - \frac{{\gamma {{\rm{K}}_0}}}{3}\frac{{{2^{ - 2l}}\Gamma (2l + 1)}}{{\Gamma (l + 2)}}\frac{{{\eta _e}}}{{w_e^3}},\\ \\
{f_{3e}} =  { - \frac{{2\delta _2^2}}{{{\pi ^2}{\rm K}_0^4}}\psi _e^9 + \frac{{\gamma {\delta _2}}}{{{\pi ^2}{\rm K}_0^2}}\frac{{{2^{ - 2l }}\Gamma (2l+1)}}{{\Gamma (l+1)}}\frac{{\psi _e^7}}{{w_e^2}}},\\ \\
{A_{11}} =  - \frac{{20}}{{{\pi ^2}}}\delta _2^2\eta _e^4 + \frac{{\gamma {\delta _2}{{\rm K}_0}}}{{{\pi ^2}}}\frac{{{2^{ - 2l + 3}}\Gamma (2l+1)}}{{\Gamma (l+1)}}\frac{{\eta _e^3}}{{w_e^2}},\\ \\
{A_{12}} =  - \frac{{\gamma {\delta _2}{{\rm K}_0}}}{{{\pi ^2}}}\frac{{{2^{ - 2l + 2}}\Gamma (2l+1)}}{{\Gamma (l+1)}}\frac{{\eta _e^4}}{{w_e^3}},\\ \\
{A_{21}} =  - \frac{{\gamma {{\rm{K}}_0}}}{3}\frac{{{2^{ - 2l}}\Gamma (2l + 1)}}{{\Gamma (l + 2)}}\frac{1}{{w_e^3}},\\ \\
{A_{22}} =  - 1 - \frac{3}{{w_e^4}} + \gamma {{\rm{K}}_0}\frac{{{2^{ - 2l}}\Gamma (2l + 1)}}{{\Gamma (l + 2)}}\frac{{{\eta _e}}}{{w_e^4}}, \\ \\
A_{23}  =  - \frac{\gamma }{3}\frac{{{2^{ - 2l + 1}}\Gamma (2l + 1)}}{{\Gamma (l + 2)}}\frac{{{\psi _e}}}{{w_e^3}}, \\ \\
{A_{32}} = { - \frac{{2\gamma {\delta _2}}}{{{\pi ^2}{\rm K}_0^2}}\frac{{{2^{ - 2l }}\Gamma (2l+1)}}{{\Gamma (l+1)}}\frac{{\psi _e^7}}{{w_e^3}}}, \\ \\
{A_{33}} = { - \frac{{18\delta _2^2}}{{{\pi ^2}{\rm K}_0^4}}\psi _e^8 + \frac{{7\gamma {\delta _2}}}{{{\pi ^2}{\rm K}_0^2}}\frac{{{2^{ - 2l }}\Gamma (2l+1)}}{{\Gamma (l+1)}}\frac{{\psi _e^6}}{{w_e^2}}}.
\end{array}\]

Next step is to diagonalize the coefficient matrix $A$ in Eq.\
\ref{coupledetaw}, so that three equation becomes uncoupled. For this purpose,
we consider the following transformation:
\begin{equation}
    \left[ {\begin{array}{*{20}{c}}
{\Delta \eta }\\
{\Delta w} \\
{\Delta \psi_0}
\end{array}} \right] = {\mathcal{P}_e}\left[ {\begin{array}{*{20}{c}}
{\Delta \tilde \eta }\\
{\Delta \tilde w} \\
{\Delta \tilde \psi_0}
\end{array}} \right],
\end{equation}
where the  matrix elements of $\mathcal{P}_e$ is chosen such that
$\mathcal{P}_e^{-1}A\mathcal{P}_e$ is diagonal. Following the standard matrix
diagonalization procedure, we obtain the nonzero elements $P_{ij}$ of
$\mathcal{P}_e$:
\[\begin{array}{l}
{P_{11}} = -{f_0}\left(  \alpha_1 - \alpha_2  - \sqrt{\Delta _0} \right),\\
{P_{12}} =  - {f_0}\left( \alpha_1 - \alpha_2  + \sqrt{\Delta _0} \right),\\
{P_{32}} = {f_1}\left(  \alpha_3 + \alpha_4  + \sqrt{\Delta _1} \right), \\
{P_{33}} = {f_1}\left(  \alpha_3 + \alpha_4  - \sqrt{\Delta _1} \right), \\
{P_{21}} =P_{22}= P_{23}=1,\\
\end{array}\]
where the constants $f_0,f_1,\alpha_1,\alpha_2,\alpha_3,\alpha_4$ and
$\Delta_0,\Delta_1$ are given by:
\[\begin{array}{l}
{f_0} = {\left( {2{\pi ^2}\gamma {{\rm{K}}_0}{g_2}w_e^{}} \right)^{ - 1}},\\
{f_1} = {\left( {24{\delta _2}{g_1}\gamma {\rm{K}}_0^2\psi _e^7{w_e}} \right)^{ - 1}},\\
{\alpha _1} = 3{\pi ^2}\left( {w_e^4 + 3} \right){2^{2l}},\\
{\alpha _2} = 3{\eta _e}\left[ {5\delta _2^2{4^{l + 1}}\eta _e^3w_e^4 + \gamma {{\rm{K}}_0}\left( {{\pi ^2}{g_2} - 16{\delta _2}{g_1}\eta _e^2w_e^2} \right)} \right],\\
{\alpha _3} = 3{\delta _2}\left( {14\gamma {\rm{K}}_0^2{g_1} - 9\delta _2^{}{2^{2l + 1}}\psi _e^2w_e^2} \right)\psi _e^6w_e^2,\\
{\alpha _4} = 3{\pi ^2}{\rm{K}}_0^4\left[ {{4^l}\left( {w_e^4 + 3} \right) - \gamma {g_2}\psi _e^2} \right],\\
\\
{\Delta _0} =  - 72\delta _2^2\left[ \begin{array}{l}
5{\pi ^2}{4^l}\left\{ {{4^l}\left( {w_e^4 + 3} \right) - \gamma {{\rm{K}}_0}{g_2}{\eta _e}} \right\}\\
 - 32{\gamma ^2}{\rm{K}}_0^2g_1^2\eta _e^2
\end{array} \right]\eta _e^4w_e^4\\
{\rm{    }} + 45\delta _2^3{2^{2l + 4}}\left( {5{\delta _2}{2^{2l}}{\eta _e}w_e^2 - 8\gamma {{\rm{K}}_0}{g_1}} \right)\eta _e^7w_e^6\\
{\rm{    }} + 96{\pi ^2}{\delta _2}\gamma {g_1}\left[ {\;{4^l}3\left( {w_e^4 + 3} \right) - 2\gamma {g_2}{{\rm{K}}_0}{\eta _e}} \right]\eta _e^3w_e^2\\
{\rm{    }} + 9{\pi ^4}{\left[ {{4^l}\left( {w_e^4 + 3} \right) - \gamma {{\rm{K}}_0}{g_2}{\eta _e}} \right]^2},\\
\\
{\Delta _1} = 81\delta _2^3\left[ {9\delta _2^{}{2^{2l}}\psi _e^2w_e^2 - 14\gamma {\rm{K}}_0^2{g_1}} \right]{2^{2l + 2}}\psi _e^{14}w_e^6\\
{\rm{   }} + 9{\pi ^4}{\rm{K}}_0^8{\left( {{4^l}\left( {w_e^4 + 3} \right) - \gamma {g_2}\psi _e^2} \right)^2}\\
{\rm{   }} + 12{\pi ^2}{\delta _2}\gamma {\rm{K}}_0^6{g_1}\left[ {21\;\left( {w_e^4 + 3} \right){4^l} - 13\gamma {g_2}\psi _e^2} \right]\psi _e^6w_e^2\\
{\rm{   }} + 36\delta _2^2{\rm{K}}_0^4\left[ {49{\gamma ^2}g_1^2\psi _e^4 + 9{\pi ^2}\gamma {g_2}{4^l}\psi _e^2 - 9{\pi ^2}\left( {w_e^4 + 3} \right){4^{2l}}} \right] \\ \times \psi _e^8w_e^4.
\end{array}\]
Also, $g_1=\Gamma(2l+1)/(2\Gamma(l+1))$ and $g_2=\Gamma(2l+1)/\Gamma(l+2)$.
After diagonalization, Eq.~\ref{coupledetaw} reduces to:
\begin{equation} \label{uncoupledetaw}
   \frac{{{d^2}}}{{d{Z^2}}}\left[ {\begin{array}{*{20}{c}}
{\Delta \tilde{\eta} }\\
{\Delta \tilde{w}} \\
{\Delta \tilde{\psi}_0}
\end{array}} \right] = \left[ {\begin{array}{*{20}{c}}
{{F_{1}}}\\
{{F_2}} \\
F_3
\end{array}} \right] + \left[ {\begin{array}{*{20}{c}}
{{-\omega_1^2}}&{{0}}& 0\\
{{0}}&{{-\omega_2^2}}& 0 \\
0 & 0 & -\omega_3^2
\end{array}} \right]\left[ {\begin{array}{*{20}{c}}
{\Delta \tilde{\eta} }\\
{\Delta \tilde{w}} \\
\Delta \tilde{\psi}_0
\end{array}} \right],
\end{equation}
where $[F_1,F_2,F_3]^{\rm T}= \mathcal{P}_e^{-1} [f_{1e},f_{2e}, f_{3e}]^{\rm
T} $. This matrix equation represents three uncoupled forced oscillators whose
solutions are  of the form
\begin{equation}
   \label{Deletawsol}
\left[ {\begin{array}{*{20}{c}}
{\Delta \tilde{\eta}  \left( Z \right)}\\
{\Delta \tilde{w}\left( Z \right)} \\
\Delta \tilde{\psi}_0(Z)
\end{array}} \right] =   \left[ {\begin{array}{*{20}{c}}
{\Delta \tilde \eta (0)\cos \left( {{\omega _1}Z} \right) + \frac{{{F_1}}}{{\omega _1^2}}\left( {1 - \cos \left( {{\omega _1}Z} \right)} \right)}\\
{\Delta \tilde w(0)\cos \left( {{\omega _2}Z} \right) + \frac{{{F_2}}}{{\omega _2^2}}\left( {1 - \cos \left( {{\omega _2}Z} \right)} \right)} \\
{\Delta \tilde \psi_0(0)\cos \left( {{\omega _3}Z} \right) + \frac{{{F_3}}}{{\omega _3^2}}\left( {1 - \cos \left( {{\omega _3}Z} \right)} \right)}
\end{array}} \right] ,
\end{equation}
where  the frequencies $\omega_1$, $\omega_2$, and $\omega_3$ are
\[\begin{array}{l}
\omega _1^2 = {\theta _0}\left( { - {\theta _1} + \;{\theta _2} + \sqrt \Omega_0  } \right),\\
\omega _2^2 = {\theta _0}\left( { - {\theta _1} + \;{\theta _2} - \sqrt \Omega_0  } \right), \\
\omega _3^2 = {\theta'}\left( {  {\theta _3} + \;{\theta _4} - \sqrt \Omega_1  } \right).
\end{array}\]

In the preceding expression,  $\theta_0, \theta',  \theta_1,\theta_2, \theta_3,
\theta_4$ are functions of the equilibrium parameters:
\[\begin{array}{l}
{\theta _0} = \frac{{{2^{ - 2l - 1}}}}{{3{\pi ^2}w_e^4}},\\
\\
\theta ' = \frac{{{2^{ - 2l - 1}}}}{{3{\pi ^2}{\rm{K}}_0^4w_e^4}},\\
\\
{\theta _1} = 3\gamma {{\rm{K}}_0}{\eta _e}\left( {16{\delta _2}{g_1}\eta _e^2w_e^2 + {\pi ^2}{g_2}} \right),\\
\\
{\theta _2} = \;{2^{2l}}\left[ {60\delta _2^2\eta _e^4w_e^4 + 3{\pi ^2}\left( {w_e^4 + 3} \right)} \right],\\
\\
{\theta _3} = {\rm{3}}\gamma {\rm{K}}_0^2\psi _e^2\left[ {{\pi ^{\rm{2}}}{\rm{K}}_0^2{g_2} - 14{\delta _{\rm{2}}}{g_1}\psi _e^4w_e^2} \right],\\
\\
{\theta _4} = 3\;\left[ {18\delta _2^2\psi _e^8w_e^4 + {\pi ^2}{\rm{K}}_0^4\left( {w_e^4 + 3} \right)} \right]{4^l}.\\
\end{array}\]

The remaining real constants $\Omega_0$ and $\Omega_1$ are:
\[\begin{array}{l}
{\Omega _0} = 3{\gamma ^2}\eta _e^2{\rm{K}}_0^2\left[ \begin{array}{l}
64{\delta _2}{g_1}\left( {12{\delta _2}{g_1}\eta _e^2w_e^2 - {\pi ^2}{g_2}} \right)\eta _e^2w_e^2\\
 + 3{\pi ^4}g_2^2
\end{array} \right]\\
\\
{\rm{        }} + 9\left[ {{\pi ^2}\left( {w_e^4 + 3} \right) - 20\delta _2^2\eta _e^4w_e^4} \right]\\
\\
{\rm{         }} \times \left[ \begin{array}{l}
{2^{4l}}\left( {{\pi ^2}\left( {w_e^4 + 3} \right) - 20\delta _2^2\eta _e^4w_e^4} \right)\\
 - \gamma {{\rm{K}}_0}{2^{2l + 1}}{\eta _e}\left( {{\pi ^2}{g_2} - 16{g_1}{\delta _2}\eta _e^2w_e^2} \right)
\end{array} \right],\\
\\
\\
{\Omega _1} = {\rm{9}}\;{\left[ {{\pi ^{\rm{2}}}{\rm{K}}_0^4\left( {w_e^4{\rm{ + 3}}} \right){\rm{ - 18}}\delta _{\rm{2}}^{\rm{2}}\psi _e^8w_e^4} \right]^{\rm{2}}}{4^{2l}}\\ \\
{\rm{        + 3}}\gamma {\rm{K}}_0^4\psi _e^4\left[ \begin{array}{l}
4\delta _{\rm{2}}^{}{g_{\rm{1}}}\left[ {147\delta _{\rm{2}}^{}g_1^{}\psi _e^4w_e^2{\rm{ - 13}}{\pi ^{\rm{2}}}\gamma {\rm{K}}_0^2{g_{\rm{2}}}} \right]\psi _e^4w_e^2\\
{\rm{ + 3}}{\pi ^{\rm{4}}}\gamma g_2^{\rm{2}}{\rm{K}}_0^4
\end{array} \right]\\ \\
{\rm{        }} - {\rm{9}}\gamma {\rm{K}}_0^2{{\rm{2}}^{{\rm{2}}l{\rm{ + 1}}}}\psi _e^2\left[ {{\pi ^{\rm{2}}}{\rm{K}}_0^4\left( {w_e^4{\rm{ + 3}}} \right){\rm{ - 18}}\delta _{\rm{2}}^{\rm{2}}\psi _e^8w_e^4} \right]\\ \\
{\rm{        }} \times \left( {{\pi ^{\rm{2}}}{\rm{K}}_0^2{g_2}{\rm{ - 14}}\delta _{\rm{2}}^{}{g_1}\psi _e^4w_e^2} \right).
\end{array}\]

\newpage


\end{document}